\documentclass[aps,prl,reprint,superscriptaddress]{revtex4-1}
\usepackage{amsmath, amssymb}
\usepackage{color}
\usepackage{graphicx}
\usepackage{CJK}

\newcommand{\ket}[1]{{| {#1}  \rangle}}
\newcommand{\bra}[1]{{\langle {#1} |}}

\newcommand{\braket}[2]{{\langle {#1} | {#2} \rangle}}

\begin{document}
\begin{CJK*}{UTF8}{} 

\title{Efficient quantum error correction of dephasing induced by a common fluctuator}

\author{David Layden}
\affiliation{Research Laboratory of Electronics, Massachusetts Institute of Technology, Cambridge, Massachusetts 02139, USA}
\affiliation{Department of Nuclear Science and Engineering, Massachusetts Institute of Technology, Cambridge, Massachusetts 02139, USA}

\author{Mo Chen \CJKfamily{gbsn}(陈墨) }
\affiliation{Research Laboratory of Electronics, Massachusetts Institute of Technology, Cambridge, Massachusetts 02139, USA}
\affiliation{Department of Mechanical Engineering, Massachusetts Institute of Technology, Cambridge, Massachusetts 02139, USA}

\author{Paola Cappellaro}
\affiliation{Research Laboratory of Electronics, Massachusetts Institute of Technology, Cambridge, Massachusetts 02139, USA}
\affiliation{Department of Nuclear Science and Engineering, Massachusetts Institute of Technology, Cambridge, Massachusetts 02139, USA}

\begin{abstract}
Quantum error correction is expected to be essential in large-scale quantum technologies. However, the substantial overhead of qubits it requires is thought to greatly limit its utility in smaller, near-term devices. Here we introduce a new family of special-purpose quantum error-correcting codes that offer an exponential reduction in overhead compared to the usual repetition code. They are tailored for a common and important source of decoherence in current experiments, whereby a register of qubits is subject to phase noise through coupling to a common fluctuator, such as a resonator or a spin defect. The smallest instance encodes one logical qubit into two physical qubits, and corrects decoherence to leading-order using a constant number of one- and two-qubit operations. More generally, while the repetition code on $n$ qubits corrects errors to order $t^{O(n)}$, with  $t$ the time  between recoveries, our codes correct to order $t^{O(2^n)}$. Moreover, they are robust to model imperfections in small- and intermediate-scale devices, where they already provide substantial gains in error suppression. As a result, these hardware-efficient codes open a potential avenue for useful quantum error correction in near-term, pre-fault tolerant devices.
\end{abstract}

\maketitle
\end{CJK*}

Decoherence, the uncontrolled decay of coherence in open quantum systems, is a central obstacle to developing coherent quantum technologies such as quantum sensors, networks, and computers. This obstacle is compounded by the destructive nature of quantum measurement: straightforward attempts to identify---and ultimately reverse---decoherence destroy the quantum coherence they seek to protect. Quantum error correction (QEC) is a technique for taming decoherence which sidesteps this issue. It encodes lower-dimensional quantum states into a higher-dimensional quantum system such that decoherence can be detected and approximately reversed without collapsing the encoded state. Specifically, the most common approach encodes $k$ logical qubits into an $n$-qubit register ($k<n$) whose Hilbert space $\mathcal{H}$ is decomposed into orthogonal subspaces $\mathcal{C}_0, \mathcal{C}_1, \mathcal{C}_2, \dots$ of dimension $2^k$ \footnote{In general, there could also be a ``remainder'' subspace $\mathcal{C}_\textsc{r}$ of arbitrary dimension so that $\mathcal{H}=\left( \bigoplus_i \mathcal{C}_i \right) \oplus \mathcal{C}_\textsc{r}$}. These subspaces are chosen by specifying operators $E_1, E_2, \dots$ and demanding that the logical states, which reside in $\mathcal{C}_0$, be mapped to $\mathcal{C}_i$ by $E_i$ without distortion \footnote{While it is possible for multiple $E_i$'s to have the same effect on the logical states, thus reducing the number of subspaces required for QEC, we will not deal with such degenerate codes here.}. By performing a partial measurement that reveals \textit{only} which subspace contains the state, and feeding back appropriately, one can reverse the occurrence of any $E_i$---and more generally, any error in $\mathcal{E} =  \text{span} \{I, E_1, E_2, \dots \}.$ The conventional strategy is to pick $E_i$'s so that $\mathcal{E}$ encompasses a broad family of operators on $\mathcal{H}$. Using Pauli operators of weight up to $w$, for instance, produces a QEC code that corrects arbitrary errors on $w$ qubits. This is a powerful approach, especially in large devices ($n \gg 1$), since it can reverse decoherence with little regard to its physical origins~\cite{nielsen, lidar}. For smaller devices, however, casting such a wide net requires an overhead of qubits ($n-k$) that is often prohibitive for near-term applications. 
A more economical strategy for small- and intermediate-scale devices is instead to use a QEC code with $\mathcal{E}$ tailored to include only the dominant, well-characterized decoherence modes. However, while this strategy is well-known (see~\cite{nielsen} \S 10.6.4), few explicit such codes have been discovered; see, e.g., Refs.\ \cite{leung:1997,robertson:2017, layden:2019}.

In order to systematically find noise-tailored QEC codes, here we focus on dephasing, since it is the dominant type of decoherence in various experiments. In particular, we consider the common scenario where dephasing in a register of qubits arises primarily due to eigenstate-preserving coupling of each qubit to a common fluctuator, which in turn exchanges energy with an external environment. That is, we consider a Hamiltonian
\begin{equation}
H = H_f^0 + \frac{1}{2}\sum_{j=1}^n \omega_j Z_j + H_f^\text{int} \otimes \sum_{j=1}^n g_j Z_j
\label{eq:H}
\end{equation}
where $[H_f^0, H_f^\text{int}]=0$, and a fluctuator that jumps incoherently between energy eigenstates $\{ \ket{\ell}_f \}$ (reflected by a dissipative term in the overall master equation). Moving to the interaction picture, the Hamiltonian \eqref{eq:H} becomes
\begin{equation}
\tilde{H} = \sum_\ell \lambda_\ell \ket{\ell}\!\bra{\ell}_f \otimes H_E,
\label{eq:H_tilde}
\end{equation}
where $H_f^\text{int} = \sum_\ell \lambda_\ell \ket{\ell}\!\bra{\ell}_f$ and $H_E := \sum_{j=1}^n g_j Z_j$. When the fluctuator is in state $\ket{\ell}_f$, qubit $j$ has an effective Hamiltonian $\lambda_\ell \, g_j  Z_j$ in the rotating frame. Jumps of the fluctuator therefore induce spatially-correlated random telegraph noise in the register, which causes dephasing~\cite{machlup:1956, neuenhan:2009}. This model, which we call \textit{common-fluctuator dephasing} (CFD), often describes the main decoherence mechanism in nuclear spins near spin defects  (e.g., Nitrogen-Vacancy centers in diamond~\cite{chen:2018}) or quantum dots, and can also be significant in superconducting qubits dispersively coupled to a common resonator with non-zero effective temperature \cite{coupling, maurer:2012, shim:2013, zaiser:2016, chen:2018,bertet:2005,bertet:2005b, gambetta:2006, majer:2007, clerk:2007, sears:2012, yan:2016, yeh:2017,yan:2018, wang:2019}. Often the register is read out and/or initialized via the fluctuator, imposing a lower limit on the desirable coupling strengths $g_j$, and making CFD a significant decoherence mode. Note that CFD does not generally produce a decoherence-free subspace (DFS).

The standard QEC approach to correct dephasing uses $E_i$'s comprising Pauli $Z$ operators on at most $w$ qubits (and $I$ on the rest). There are $\sum_{m = 0}^w \binom{n}{m}$ such matrices; a simple counting argument (the quantum Hamming bound applied to phase noise) therefore suggests that $n \ge 2w+1$ physical qubits are required to protect $k=1$ logical qubit from arbitrary phase errors of weight $\le w$~\cite{nielsen}. Indeed, the repetition code saturates this bound: the smallest instance uses $n=3$ for $w=1$, has logical states $\ket{0_\textsc{l}}=\ket{+ \! + \!+}$ and $\ket{1_\textsc{l}}=\ket{- \!- \!-}$ where $\ket{\pm} := \frac{1}{\sqrt{2}} (\ket{0}\pm \ket{1})$, and corrects for $\mathcal{E} = \text{span} \{I, Z_1, Z_2, Z_3\}$. It can correct CFD as follows: In any run of the experiment, the register evolves over time $t$ as $U(\theta)=e^{-i \theta H_E}$ for some random variable $\theta \in [t \lambda_\text{min}, t \lambda_\text{max}]$ that depends on the fluctuator's trajectory. For short $t$ (understood in units of $1/\max_{j \ell}|g_j \lambda_\ell|$, and often reducible through dynamical decoupling~\cite{viola:1998, ban:1998, biercuk:2009, chen:2018}), $U(\theta)$ can be approximated as $U(\theta) = I - i \theta H_E + O(t^2)$.
Since $\theta H_E \in \mathcal{E}$ regardless of $\theta$, this 3-qubit code corrects dephasing at order $O(t)$. More generally, $H_E^q$ contains Paulis of weight $\le q$, so correcting to order $O(t^q)$ with the repetition code requires $n=2q+1$ qubits (for $k=1$). 

While the value of $\theta$ is unknown and varies from one run to the next, the coupling strengths $g_j$ are often fixed and well characterized. This suggests designing a code that corrects expressly for $\mathcal{E} = \text{span} \{ I, H_E, H_E^2, \dots, H_E^q \}$, and depends on the $\{ g_j \}$ in a particular device. A similar counting argument as above suggests that such a code would require $q+1$ subspaces to protect a logical qubit to order $O(t^q)$, and therefore require 
\begin{equation}
n= \lceil 1+\log_2 (q+1) \rceil
\label{eq:n_vs_l}
\end{equation}
qubits---an exponentially smaller overhead. We give a family of such codes here for general $q$ and arbitrary coupling strengths $\{g_j \}$. We focus in particular on the $q=1$ case, where one logical qubit is encoded in two physical qubits rather than three. We construct recovery and logical operations for this code, which can be implemented using a constant number of one- and two-qubit operations.

\begin{figure}
    \centering
    \includegraphics[width=0.48\textwidth]{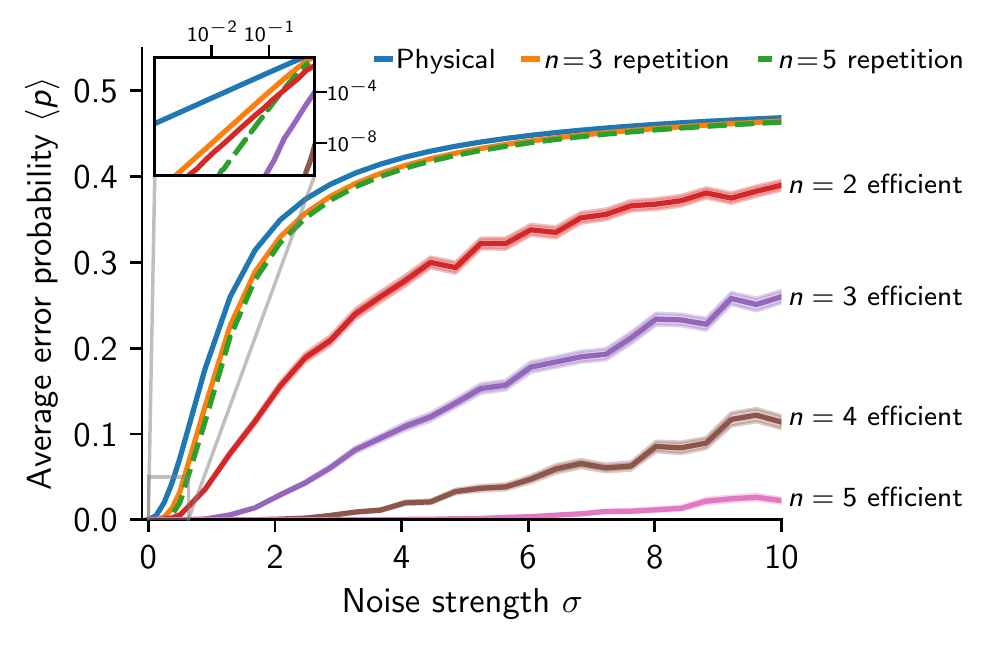}
    \caption{\textbf{Comparison of QEC codes performance.}
We assume that the effect of the quantum fluctuator is to impart a random phase, $\theta$, which follows a Gaussian distribution $\theta \sim \mathcal{N}(0,\sigma)$ with standard deviation $\sigma$. By normalizing the $g_j$'s to lie in $[0,1]^n$, $\sigma$  describes the noise strength. 
CFD followed by a QEC recovery (if applicable) results in an effective phase- or bit-flip channel $\rho \mapsto (1-p) \rho + p\, A \rho A$, where $A=Z$ for the physical qubits, $X_\textsc{l}$ for the repetition codes, and $Z_\textsc{l}$ for hardware-efficient codes. The average infidelity,  average trace distance and  diamond distance to $I$ are all $\propto p$. As the performance of all strategies shown depends on $\{ g_j \}$,  we plot the average of $p$ over $\{g_j\} \in [0,1]^n$. The error bands for the hardware-efficient codes denote the standard error of the mean from Monte Carlo integration. More details on the numerical implementation are given in~\cite{SM}.}
    \label{fig:p_vs_sigma}
\end{figure}

The decomposition $\mathcal{H}$ into subspaces $\mathcal{C}_i$ for QEC is equivalent to the Knill-Laflamme conditions~\cite{knill:1997, beny:2011}. For $k=1$ and $\mathcal{E}=\text{span}\{H_E^j\}_{j=0}^q$, these take the form
\begin{gather}
\bra{0_\textsc{l}} H_E^m \ket{0_\textsc{l}}
=
\bra{1_\textsc{l}} H_E^m \ket{1_\textsc{l}} \label{eq:KL1}\\
\bra{0_\textsc{l}} H_E^m \ket{1_\textsc{l}} = 0 \label{eq:KL2}
\end{gather}
for $0 \le m \le 2q$, where we consider values of $q$ that saturate the ceiling in Eq.~\eqref{eq:n_vs_l} (that is, $q = 2^{n-1}-1$). Finding a QEC code that corrects this $\mathcal{E}$ therefore requires finding logical states $\ket{0_\textsc{l}}$ and $\ket{1_\textsc{l}}$ that satisfy Eqs.~\eqref{eq:KL1} and \eqref{eq:KL2}. We begin with the ansatz
\begin{equation}
\ket{0_\textsc{l}} = \sum_{j=0}^{2^n-1} r_j e^{i \theta_j} \ket{j}
\qquad
\ket{1_\textsc{l}} = \sum_{j=0}^{2^n-1} r_{(2^n-1-j)} e^{i \phi_j} \ket{j},
\label{eq:ansatz}
\end{equation}
for $r_j, \theta_j, \phi_j \in \mathbb{R}$, where we use $\ket{j}$ to denote the $n$-bit binary representation of the integer $j$. That is, we fix the amplitudes of $\ket{1_\textsc{l}}$ to be those of $\ket{0_\textsc{l}}$ in reverse order. Notice that Eq.~\eqref{eq:ansatz} always satisfies \eqref{eq:KL1} for even $m\ge 0$, since $X^{\otimes n} H_E^m X^{\otimes n} = (-1)^m H_E^m$. For odd $m$:
\begin{equation}
\bra{0_\textsc{l}} H_E^m \ket{0_\textsc{l}} = - \bra{1_\textsc{l}} H_E^m \ket{1_\textsc{l}} = \vec{z} \cdot \vec{v}_m,
\end{equation}
where $\vec{z}, \, \vec{v}_m \in \mathbb{R}^{q + 1}$  are defined as $z_i = \bra{i} Z_\textsc{l} \ket{i}$, with $Z_\textsc{l} := \ket{0_\textsc{l}} \! \bra{0_\textsc{l}} - \ket{1_\textsc{l}} \! \bra{1_\textsc{l}}$, and $(\vec{v}_m)_i = \bra{i} H_E^m \ket{i}$ for $i\in[0,q]$ and odd $m \in [0,2q]$. Therefore, Eq.~\eqref{eq:KL1} is satisfied for all relevant $m$ if $\vec{z} \perp \text{span} \{ \vec{v}_m \}$. We can always find such a $\vec{z}$ ($\neq \vec{0}$) since the $\vec{v}_m$'s have dimension $q + 1$ but there are only $q$ of them, so they cannot form a complete basis. One approach is to construct a matrix $V$ with $\vec{v}_m$'s as columns; then, $I-VV^+$ projects onto $\text{span} \{ \vec{v}_m \}^\perp$ (where $+$ and $\perp$ denote the pseudoinverse and orthogonal complement, respectively) and therefore has at least one real eigenvector $\vec{u}$ with unit eigenvalue \footnote{Alternatively, the modified Gram-Schmidt procedure provides a less intuitive but more numerically stable method.}. Taking $\vec{z} = \vec{u}/||\vec{u}||_1$ satisfies Eq.~\eqref{eq:KL1} since $\vec{u} \cdot \vec{v}_m=0$ automatically. Finally, building upon a technique developed in Ref.\ \cite{layden:2019} for optimization, we pick $r_j$'s as
\begin{equation}
(r_j, r_{(2^n-1-j)})
=
\begin{cases}
(0, \sqrt{z_j}), & \text{if } z_j \ge 0\\
(\sqrt{-z_j},0), & \text{if } z_j < 0.
\end{cases}
\label{eq:rj}
\end{equation}
This choice ensures that $\braket{j}{0_\textsc{l}}$ or $\braket{j}{1_\textsc{l}}$ vanishes for every $j$, thus satisfying Eq.~\eqref{eq:KL2}. We now have normalized logical states that form a valid QEC code for all $q \ge 1$. Notice that the components of $\ket{0_\textsc{l}}$ and $\ket{1_\textsc{l}}$ generically have unequal amplitudes $r_j$ by necessity, in marked contrast with classical error-correcting codes and most known QEC codes. The phases $\theta_j$ and $\phi_j$ can be chosen arbitrarily---we demonstrate a convenient choice below. The performance of these codes on $n \le 5$ qubits is shown in Fig.~\ref{fig:p_vs_sigma} using an illustrative model of a normally-distributed $\theta$. In addition, we give the pseudothresholds for $n=2$ and 3 under the same model in the Supplemental Material~\cite{SM}.

To illustrate this QEC code, we consider explicitly the smallest case of $n=2$ qubits coupled to a two-level fluctuator with $\lambda_{\pm 1} = \pm 1$ [cf.\ Eq.~\eqref{eq:H_tilde}],  at high temperature. We will label the register qubits 1 and 2 such that $|g_1| \ge |g_2|$. Note that here---and in general---$H_E=g_1 Z_1 + g_2 Z_2$ is a combination of weight-1 Pauli operators, not a weight-2 Pauli. This $H_E$ gives $\vec{v}_1 = (g_1+g_2, g_1-g_2)^\top$. The matrix $I-VV^+$ has only a 1-dimensional eigenspace with unit eigenvalue, spanned by $\vec{u}=(-g_1+g_2, g_1+g_2)^\top$, where $\vec{u} \cdot \vec{v}_1 = 0$. If $g_1 > 0$ we find $r_1 = r_3 =0$ and
\begin{equation}
r_0 = c \sqrt{g_1-g_2} \qquad r_2 = c\sqrt{g_1+g_2},
\end{equation}
where $c = 1/\sqrt{||\vec{u}||_1}$. This gives logical states
\begin{equation}
\ket{0_\textsc{l}} = \ket{\chi_0} \ket{0}
\qquad
\ket{1_\textsc{l}} = \ket{\chi_1} \ket{1}
\label{eq:codewords}
\end{equation}
with
\begin{equation}
\begin{split}
\ket{\chi_0} &= c \Big( \sqrt{|g_1-g_2|}\,  e^{i \theta_0} \, \ket{0} + \sqrt{|g_1+g_2|} \, e^{i\theta_2} \, \ket{1} \Big)\\
\ket{\chi_1} &= c \Big( \sqrt{|g_1+g_2|}\,  e^{i \phi_1} \, \ket{0} + \sqrt{|g_1-g_2|} \, e^{i\phi_3} \, \ket{1} \Big),
\end{split}
\label{eq:chi}
\end{equation}
where $\ket{0}$ and $\ket{1}$ refer to the states of a qubit. The $g_1<0$ case gives the same result up to a relabelling of $\ket{0_\textsc{l}} \leftrightarrow \ket{1_\textsc{l}}$. This code corrects for $\mathcal{E} = \text{span}\{I, H_E \}$; by design, however, it does not correct for $Z_1 Z_2$, nor $Z_1$ or $Z_2$ individually, none of which belong to $\mathcal{E}$. Rather, it corrects CFD with fewer qubits than the smallest repetition code precisely because we have chosen not to correct individual Pauli operators. 

\begin{figure}
\includegraphics[width=0.48\textwidth]{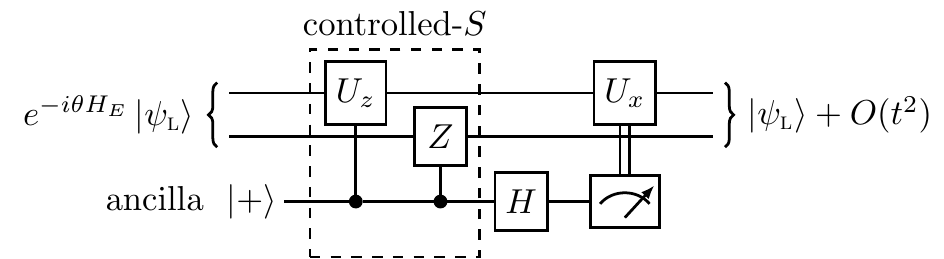}
\caption{A recovery procedure for $n=2$ qubits where $\ket{\psi_\textsc{l}} = \alpha \ket{0_\textsc{l}} + \beta \ket{1_\textsc{l}}$ for arbitrary $\alpha$ and $\beta$, $H$ denotes a Hadamard gate, and $\theta$ is a random variable. The unitaries $U_x$ and $U_z$ are both $\pi$ rotations about orthogonal axes on the Bloch sphere which are determined by $g_1$, $g_2$ and $\vartheta$.}
\label{fig:recovery}
\end{figure}

Observe that Eqs.~\eqref{eq:codewords} and \eqref{eq:chi} reduce to a DFS in the limit where one exists ($|g_1| = |g_2|$), but this is in practice rare. More generally, notice that the choice $\theta_0 = \phi_1+\pi = -\theta_2 = -\phi_3 = \vartheta$ for arbitrary $\vartheta$ proves convenient: First, it gives $\braket{\chi_0}{\chi_1}=0$, and a simple action of $H_E$ on logical states:
\begin{equation}
\begin{split}
H_E \, \ket{0_\textsc{l}} &\propto \ket{\chi_1} \ket{0} =: \ket{0_\textsc{e}}\\
H_E \, \ket{1_\textsc{l}} &\propto \ket{\chi_0} \ket{1} =: \ket{1_\textsc{e}}.
\end{split}
\label{eq:E_effect}
\end{equation}
Both lines have the same proportionality constant, and we have defined the error states $\ket{0_\textsc{e}}$ and $\ket{1_\textsc{e}}$. We emphasize that since $H_E$ cannot generically be decomposed as a tensor product, it maps most separable states to entangled states; Eq.~\eqref{eq:E_effect}---wherein the first qubit is ``flipped" by $H_E$---is due to our choice of $\ket{0_\textsc{l}}$ and $\ket{1_\textsc{l}}$. Second, consider the orthogonal projectors $P_\textsc{l} = \ket{0_\textsc{l}}\!\bra{0_\textsc{l}} + \ket{1_\textsc{l}}\!\bra{1_\textsc{l}}$ and $P_\textsc{e} = \ket{0_\textsc{e}}\!\bra{0_\textsc{e}} + \ket{1_\textsc{e}}\!\bra{1_\textsc{e}}$ onto $\mathcal{C}_0 = \text{span}\{\ket{0_\textsc{l}}, \ket{1_\textsc{l}} \}$ and $\mathcal{C}_1 = \text{span}\{\ket{0_\textsc{e}}, \ket{1_\textsc{e}} \}$ respectively ($\mathcal{H} = \mathcal{C}_0 \oplus \mathcal{C}_1$). One can detect an error non-destructively by measuring parity in the $\ket{\chi_i} \ket{j}$ basis, which can be done by performing phase estimation (i.e., ``phase kickback") on
\begin{equation}
S =
P_\textsc{l} - P_\textsc{e} = U_z \otimes Z_2
\label{eq:stabilizer}
\end{equation}
with an ancilla~\cite{cleve:1998}. Crucially, the choice of phases in $\ket{0_\textsc{l}}$ and $\ket{1_\textsc{l}}$ makes $S$ separable here, where  $U_z := \ket{\chi_0} \! \bra{\chi_0} - \ket{\chi_1} \! \bra{\chi_1}$ is a $\pi$ rotation about some axis determined by $g_1$, $g_2$ and $\vartheta$. This means that the controlled-$S$ (c$S$) operation used to measure the error syndrome can be implemented through a pair of two-qubit operations (c$U_z$ and c$Z$), rather than a more challenging 3-qubit operation. If an error is detected, it can be corrected by applying $U_x := \ket{\chi_0} \! \bra{\chi_1} + \ket{\chi_1} \! \bra{\chi_0}$ to qubit 1---a $\pi$ rotation about a different axis. (Both $U_x$ and $U_z$ could be synthesized out of a constant number of Pauli rotations, or implemented directly, e.g., by driving qubit 1 off resonance~\cite{mckay:2017}.) The full recovery procedure, which corrects CFD to leading order, is shown in Fig.\ \ref{fig:recovery}. Note that $S$ behaves like a stabilizer, in the sense of its action on $\mathcal{C}_0$ and $\mathcal{C}_1$. It does not, however, fit in the usual QEC stabilizer formalism since $\{H_E, S\} \neq 0$ generically, because $\{ H_E, S \} \! \ket{\psi} = 0$ for $\ket{\psi} \in \mathcal{C}_0$ but not for $\ket{\psi} \in \mathcal{C}_1$~\cite{gottesman:1997}. This is because $H_E$ maps $\mathcal{C}_0$ to $\mathcal{C}_1$ without distortion, but not vice-versa, as $H_E$ is not generically in the Pauli group. (Neither is $S$.) In spite of these unusual features, the procedure for feeding back on $S$ in Fig.~\ref{fig:recovery} is largely the same as that of the usual stabilizer formalism. Finally, (i) the encoding can be realized by applying a $c_2(U_x)_1$ gate to an initial state $\ket{\chi_0} \ket{\psi}$, and (ii) there is a simple way to implement any logical unitary $U_\textsc{l}$ in this code: apply the corresponding physical $U$ to qubit 2 followed by a recovery.

The logical states derived above are also valid for all $q > 1$ (i.e., $n>2$ qubits), but the corresponding recovery and logical operations are generally more involved. Generically, the analogues of $S$ in \eqref{eq:stabilizer} are not separable for any choice of $\theta_j$ and $\phi_j$ \footnote{e.g., $S_1 = P_\textsc{l} + P_{\textsc{e}\scriptscriptstyle 1} - P_{\textsc{e}\scriptscriptstyle 2} - P_{\textsc{e}\scriptscriptstyle 3}$ and $S_2 = P_\textsc{l} - P_{\textsc{e}\scriptscriptstyle 1} + P_{\textsc{e}\scriptscriptstyle 2} - P_{\textsc{e}\scriptscriptstyle 3}$, which could be measured sequentially to identify an error for $n=3$}. One might still synthesize them with one- and two-qubit operations,   perform phase kickback through optimal control, or implement a QEC recovery via more general channel-engineering techniques~\cite{khaneja:2005, defouquieres:2011, lloyd:2001, shen:2017}. 
More efficient solutions could even be found by analyzing specific experimental scenarios.
One approach could be for example to use devices with $\{g_j\}$ chosen so that the recovery and logical operations can be conveniently implemented. One could also correct to a slightly lower order $q$ [i.e., maintaining $n=O(\log q)$ but not saturating the ceiling in Eq.~\eqref{eq:n_vs_l}]; this would yield a continuous family of possible $\vec{z}\,$'s [cf.\ Eq.~\eqref{eq:rj}], among which one might find codes with convenient QEC operations.
Note finally that for $n>2$ it is not the bare $H_E^m$'s that map the codespace to the orthogonal subspaces $\{\mathcal{C}_i\}_{i\ge 1}$, but rather linear combinations of them.

These noise-adapted QEC codes involve a trade-off: they correct CFD very efficiently at the cost of leaving most other errors uncorrected. For instance, errors during gates, due to miscalibration of $g_j$'s, or from decoherence beyond CFD will generally affect the logical state~\cite{SM}. Accordingly, these codes are manifestly not fault-tolerant in their current form~\cite{nigg:2014}. Crucially though, they offer such a large error budget under strong CFD---as evidenced by the gaps between QEC codes and physical qubits in Fig.~\ref{fig:p_vs_sigma}---that this trade-off can easily be worthwhile, much like the targeted correction of photon loss in~\cite{ofek:2016}. Indeed, as we show in~\cite{SM}, the gap survives even in the presence of large miscalibration of the $g_j$'s. Fault-tolerance could still be achieved  using implementation-specific methods as in Ref.\ \cite{rosenblum:2018}.
In the long-term, concatenation could potentially reach fault-tolerance, using our noise-adapted codes at the lowest level of encoding to protect against the dominant error source, and more conventional codes at higher levels. Even more importantly, our codes  could have a near-term impact in applications such as quantum sensing and communication, where long-lived quantum memories are useful even when they are not fault-tolerant. We emphasize, however, that these codes are designed expressly for small- and medium-scale qubit registers, and that the exponential reduction in overhead should be understood to apply only in such devices. For one, there is typically a maximum $n$ above which CFD no longer dominates. Also, while the error budget always increases with $n$ in principle, so too do the effects of gate errors, miscalibration of $g_j$'s and decoherence beyond CFD, as more qubits introduce more error channels. Conversely, this growing sensitivity suggests an unconventional quantum sensing scheme to measure $\{g_j\}$ for large $n$, by variationally adjusting one's estimates to maximize code performance. In the nearer term, however, these imperfections will likely set a maximum $n$ in any particular device beyond which one achieves no further gains, depending on their relative importance compared to CFD~\cite{SM}.

The QEC codes presented could be generalized in several ways. First, they can readily be made to correct dephasing due to multiple common fluctuators given enough qubits, at the cost of correcting to lower order in $t$. Similarly, they can correct spatially-correlated phase noise beyond that arising from common fluctuators. For instance, classical white noise in the energy gaps of register qubits leads to Lindblad error operators $L_j = \sqrt{\lambda_j} \, \vec{c}_j \cdot (Z_1, \dots, Z_n)$, where $\{ \sqrt{\lambda_j} \, \vec{c}_j \}$ describes the noise's normal modes~\cite{layden:2018}. In the limit of spatially uncorrelated noise the $L_j$'s become Pauli $Z$ operators; however, correlated noise produces $L_j$'s with unequal amplitudes $\sqrt{\lambda_j}$. When the noise correlations are appreciable, it could be advantageous to use a QEC code that corrects the stronger noise modes (those with large $\lambda_j$'s) to higher order in $t$ than the weaker ones (smaller $\lambda_j$'s) through an appropriate choice of $V$. It may also be possible to extend the codes presented here for the setting where a fluctuator's state affects not only the energy gap of each qubit, but also the direction of its Hamiltonian (i.e., its quantization axis)~\cite{aiello:2015}. Eigenstate-preserving coupling arises frequently in practice because a large detuning between a weakly-coupled qubit and fluctuator suppresses non-commuting parts of their interaction Hamiltonian. However, when the coupling to the fluctuator is comparable to the internal Hamiltonian, such as for nuclear spins near defects in diamond, there can remain significant non-commuting terms leading to $H_E \sim \sum_j \vec{g}_j \cdot \vec{\sigma}_j$ in Eq.~\eqref{eq:H_tilde}. We analyze this effect's impact on code performance in~\cite{SM}. Extending the codes introduced here to this more general setting would make them even more widely applicable to near-term experiments, but at the cost of larger overheads, since they would need to contend with a substantially larger space of possible errors. It may be more practical instead to suppress non-commuting interaction terms at the hardware level by increasing the energy gaps $\omega_j$ of the register qubits, or at the ``software" level through concatenation~\cite{SM}. Another interesting generalization would be to efficiently encode $k > 1$ logical qubits, which seems plausible based on the counting argument used throughout involving the dimension of $\mathcal{H}$ versus $\mathcal{E}$. Finally, it would be interesting to use the tools presented here to design codes for other common error sources, such as other types of decoherence or control/measurement errors.

Our results demonstrate that it is possible to find noise-adapted QEC codes with a well-defined advantage (here exponential) over known, general codes. It is commonly argued that QEC will be of little use in Noisy Intermediate-Scale Quantum (NISQ) devices due to its prohibitive overhead~\cite{preskill:2018}. Noise-adapted QEC codes are a promising way to reduce this overhead, although to date they have mostly relied on numerical and variational techniques that lack transparency in terms of what advantage the codes can offer, and when~\cite{reimpell:2005, fletcher:2007, kosut:2008, taghavi:2010, johnson:2017} (see also~\cite{lidar} Ch.\ 13 and~\cite{noh:2018}). In contrast, the codes introduced here exhibit a clear reduction in overhead under a well-characterized and common type of noise. New QEC codes of this type could provide a middle ground between small-scale uncorrected devices and large-scale fault-tolerant ones, where the dominant decoherence mechanisms are tamed through specialized codes with only modest overheads.  This view of near-term QEC as quantum ``firmware" rather than ``software" suggests a possible interplay between theory and experiment, whereby NISQ hardware and efficient QEC codes both guide each other's development.

\acknowledgements
We wish to thank Isaac Chuang, Liang Jiang, Morten Kjaergaard, Yi-Xiang Liu, William Oliver and Peter Shor for helpful discussions. This work was supported in part by the U.S.\ Army Research Office through MURI grant No.\ W911NF-15-1-0548, and by NSF grants EECS1702716 and EFRI-ACQUIRE 1641064.

\nocite{dreau:2012}

\bibliography{references}
\end{document}